\documentclass[prl,showpacs,superscriptaddress,twocolumn,a4paper]{revtex4-1}
\usepackage{amsmath}
\usepackage{amsfonts}
\usepackage{amssymb}
\usepackage{graphicx}
\usepackage{textcomp}
\usepackage{color}
\usepackage{array,tabularx}
\usepackage{psfrag}

\def\<{\langle}
\def\>{\rangle}

\def\Dxe{\Delta x^\cap}
\def\Dxi{\Delta x^\cup}
\def\De{D^\cap}
\def\Di{D^\cup}

\def\Px{P_{\cap}}

\def\Pu{P_{\rm d}}

\begin{document}

\title{Escape rate and diffusion of a random walker}
%\title{Escape rate and diffusion in periodic potentials: Langevin v.s. random walk}
%\title{Langevin v.s. random walk escape and diffusion}
\author{Antonio Piscitelli}\email[]{AntPs@ntu.edu.sg}
\affiliation{
Division of Physics and Applied Physics, School of Physical and Mathematical 
Sciences, Nanyang Technological University,
Singapore
}
\author{Massimo Pica Ciamarra}\email[]{massimo@ntu.edu.sg}
\affiliation{
Division of Physics and Applied Physics, School of Physical and Mathematical 
Sciences, Nanyang Technological University,
Singapore
}
\affiliation{
CNR--SPIN, Dipartimento di Scienze Fisiche, Universit\`a di Napoli Federico II, 
I-80126, Napoli, Italy
}

\date{\today}

\begin{abstract}
We determine the rate of escape from a potential well, and the diffusion coefficient in a periodic potential,
of a random walker that moves under the influence of the potential in between successive collisions with the heat bath. 
In the overdamped limit, both the escape rate and the diffusion coefficient coincide with those of a Langevin particle.
Conversely, in the underdamped limit the two dynamics have a different temperature dependence. In particular,
at low temperature the random walk has a smaller escape rate, but a larger diffusion coefficient.
\end{abstract}

\maketitle
The evaluation of the rate of escape of a particle from a one dimensional potential well,
and of the diffusion coefficient of a particle moving in a one dimensional
periodic potential, are classical problems in statistical physics
that are relevant to the physical, chemical, engineering and biological sciences.
When the timescale of interaction of the particle with the heat bath is the smallest
timescale of the problem~\cite{kubo_fluctuation-dissipation_1966},
escape and diffusion can be investigated within a Langevin formalism.
In this context, solutions  have been obtained ~\cite{hanggi_reaction-rate_1990}
in both the overdamped, $\tau_{\rm vis} \ll \tau_{\rm cross}$, and underdamped limits, $\tau_{\rm vis} \gg \tau_{\rm cross}$. 
Here $\tau_{\rm vis} = m/\gamma$ is the viscous relaxation timescale,
with $m$ mass of the particle and $\gamma$ the viscous friction coefficient, and $\tau_{\rm cross} = 1/\omega_b$ 
is a timescale related to the exchange between kinetic and potential energy during barrier crossing
fixed by the shape of the potential $V(x)$
on the top of the barrier, $\omega_{b}^2 = \frac{1}{m} \left| \partial^2 V/\partial x^2 \right|_{b}$.
In a variety of different
contexts including research strategies in biology~\cite{bartumeus_influence_2008}, transport in electronics~\cite{jacoboni_monte_1989},
market evolution models~\cite{boghosian_h_2015}, supercooled liquids~\cite{weeks_three-dimensional_2000,eaves_spatial_2009,pastore_dynamic_2015},
diffusion of atoms in optical lattices~\cite{kindermann_nonergodic_2016}, diffusion of molecules at liquid/solid interfaces~\cite{skaug_single-molecule_2014},
the time scale on interaction with the heat bath is not the smallest one,
and the Langevin approach is no longer justified.
In these cases one should adopt 
a random walk formalism, allowing the particle to
move under the influence of the potential in between successive collisions with the heat bath.

In this Letter we report on the first investigation of escape and diffusion problems
within this formalism. We consider a simple model in which a walker interacts with the heat bath with a constant rate $t_c$,
the interactions instantaneously randomizing the walker's velocity according to
the Maxwell-Boltzmann distribution at the considered temperature. When not interacting with the heat bath,
the walker moves according to Newton's equation within the potential. 
As a model potential, we have considered a periodic $x^4$ potential, $V(x)$, with period $L$ and energy barrier $V(\pm L/2) = \Delta U$,
but our results are easily generalizable. 
Thus in the period $ -L/2 \leq x \leq L/2$, the potential is $V(x) = \frac{1}{2} m\omega_0^2 x^2 - \frac{m\omega_0^2}{L^2} x^4$.
The energy barrier is $\Delta U = \omega_0^2 L^2 / 16$, while $\omega_{b}^2 = 2\omega_0^2$.
We have determined the escape rate and the diffusion coefficient in both
the overdamped, $t_c^{-1} \ll \tau_{\rm cross}$, and the underdamped limits, $t_c^{-1} \gg \tau_{\rm cross}$,
validating our theoretical results against numerical simulations. 
We show that in the overdamped limit the random walk and the Langeving dynamics
give consistent results. Conversely, the two dynamics differ in the underdamped limit,
the Langevin dynamics having an higher escape rate but, surprisingly, a smaller diffusion coefficient.

\begin{figure}[t]
\psfrag{tag1}{$\Pu$}
\psfrag{tag}{$\Px \exp(\Delta U/T)$}
\centering
 \includegraphics*[scale=0.33]{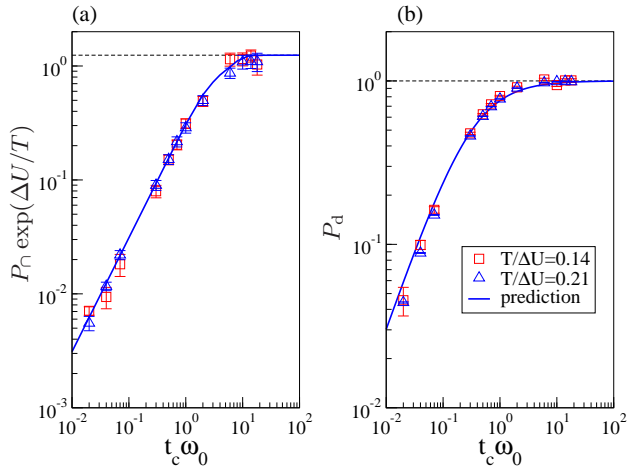} 
 \caption{Panel a: probability $\Px$ that a jumps crosses an energy barrier, normalized with the Arrhenius factor. The full line is our
 theoretical prediction, the dashed line the asymptotic value for $\Delta T/U = 0.21$. 
 Panel b: fraction of uncorrelated barrier crossing jumps, $\Pu$. The full line
 is an empirical fitting formula based on the predicted behavior in the $t_c \to 0,\infty$ limits (see text).
 }
 \label{fig:emmepf}
 \end{figure}

The escape rate is conventionally defined
as the rate with which a particle irreversibly escapes from a well in a given direction. 
The rate can be determined considering that the escaping process occurs through three uncorrelated steps,  as a particle 
first crosses the energy barrier in the considered direction, 
then recrosses the barrier an {\it even} number of times, 
and finally moves away from the top of the barrier decorrelating, without the occurrence of any further recrossing. 
We indicate the probabilities of these events with $\Px/2$, $p$ and $\Pu$, respectively,
so that $\Px$ is the probability that a thermalized particle performs a barrier crossing jump, regardless of its direction.
Accordingly, the escape rate is given by $\Gamma_{\rm RW} = \Px p \Pu/(2 t_c)$. 
Since the probability that a particle recrosses a barrier an even number of times is $p = \sum_{n=0}^{\infty} \Pu (1-\Pu)^{2n} = (2-\Pu)^{-1}$,
one finally gets
\begin{equation}
\Gamma_{RW}(t_c) = \frac{\Px(t_c)\Pu(t_c)}{2 t_c (2-\Pu(t_c))}. 
\label{eq:rate}
\end{equation}
To estimate the escape rate, we thus need to estimate the barrier crossing probability $\Px(t_c)$,
and the decorrelating probability $\Pu$. $\Px$ is simply obtained from an equilibrium average over the jumps. 
Indeed, each jump is characterized by three variables,
the coordinate of the starting point, $x_s$, the initial velocity $v_s$, and the time of flight $t$.
$x_s$ and $v_s$ have a Boltzmann and Maxwellian equilibrium distribution, respectively, 
while $t$ is exponentially distributed with time constant $t_c$. Alternatively, 
each jump can be characterized by 
$x_s$, by the coordinate of the final point, $x_e$, and by the total energy $E$.
Assuming with no loss of generality $\left| x_s \right| < L/2$, $\Px$ is the probability 
that $\left| x_e \right| > L/2$, which is found to be
$\Px =2 \int_{-L/2}^{L/2} dx_s \int_{L/2}^\infty dx_e \int_{0}^\infty dE f(x_s,x_e, E)$, where 
\begin{equation}
f= \frac{1}{{\emph Z}(T)v(x_s,E) v(x_e,E)} \frac{e^{-\frac{t_E(x_s\rightarrow x_e)}{t_c}}}{\sqrt{2\pi m T}t_c}   \frac{e^{-\frac{E}{T}}}{\sqrt{2\pi T}}.
 \label{eq:f}
\end{equation}
Here $f(x_s,x_e, E)$ is the probability the walker interacts with the heat bath when in position $x_s$, that through this interaction
it acquires a total energy $E$, and that its flight time equals the time needed to travel from $x_s$ to $x_e$ with total energy $E$,
which is given by $t_E(x_s\rightarrow x_e) =\int_{x_s}^{x_e} \frac{dz}{v(z,E)}$. 
$Z=\int_{-L/2}^{L/2} e^{-\frac{V(u)}{T}}du$ is a temperature dependent normalization constant. 
%while the velocities at $x_s$ and $x_e$ are Jacobians resulting from the transformation of variables.
The predicted dependence of $\Px$ on $t_c$ illustrated in Fig.~\ref{fig:emmepf}a (full line) correctly
describes the numerical results. The high and small $t_c$ limits can be rationalized.
In the $t_c \to 0$ limit, the above triple integral can be carried out,
and one finds $\Px = \omega_0 t_c \pi^{-1} \exp(-\Delta U/T)$. In the $t_c \to \infty$ limit all jumps 
with enough energy cross the barrier and $\Px$ approaches a constant, whose
weak temperature dependence is neglected in the following~\cite{piscitelli_preparation_2016}.
 
The decorrelation probability $\Pu$ is estimated considering that,
%without further recrossing, can be estimated in a mean
%first passage time formalism, as in the case of the Langevin dynamics~\cite{feller_introduction_1957}.
if a particle reaches a position which is at a far enough distance $l_T$ from the top of the barrier
without recrossing, then it decorrelates as its dynamics becomes dominated by the potential.
We assume $l_T$ to be the distance at which the potential significantly affects the velocity $v(x,E) = \sqrt{(2/m)(E -V(x))}$ 
of a particle crossing the barrier with energy $E$, and estimate $l_T \simeq \sqrt{T/m\omega^2_b}$
through a Taylor expansion of $v(x,E)$ around the top of an energy barrier.
In the overdamped limit $\omega_b t_c \ll 1$, $\Pu$ is given by the probability that a barrier crossing
event is followed by a sequence of jumps (of typical size $\propto \sqrt{T}$) 
able to drive the particle at distance $l_T \propto \sqrt{T}$ from the top, before a recrossing occurs.
It is easy to show in a mean first passage time formalism~\cite{feller_introduction_1957} that in this regime $\Pu(t_c) \simeq 2 \omega_b t_c$.
In the underdamped limit, $\omega_b t_c \gg 1$, jumps are long and the barrier crossing events can be considered irreversible, so that $\Pu = 1$. 
The numerical measure of $\Pu(t_c)$ confirms our predictions for the overdamped and underdamped limits of $\Pu$,
as illustrated in Fig.~\ref{fig:emmepf}b. 
We approximate in the following $\Pu$ with a simple functional form able to 
capture the crossover between the $\omega_b t_c \ll 1$ and the $\omega_b t_c \gg 1$ limits,
$\Pu(t_c) = \frac{2 \omega_0 t_c}{k + 2\omega_0 t_c}$. We fix $k = \omega_0/\omega_b$ exploiting an 
analogy with the Langevin dynamics in the overdamped low temperature limit, we will detail below.

\begin{figure}[t]
\centering
\includegraphics*[scale=0.33]{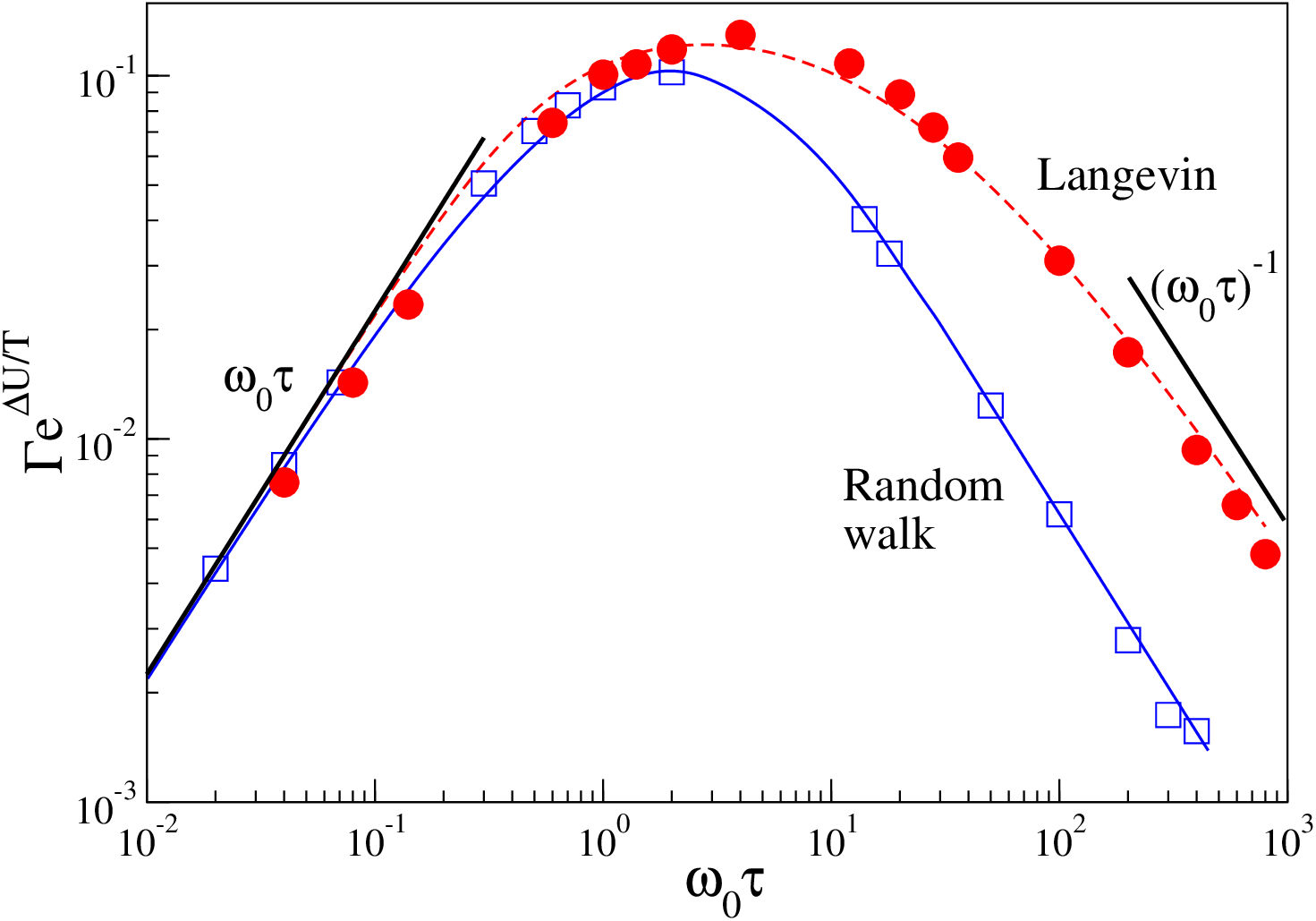}
\caption{Numerical results and theoretical predictions for dependence of the 
jump rate on $\tau = t_c = \tau_{\rm vis}$, for the Random Walk dynamics and the Langevin dynamics.
The temperature is $T /\Delta U= 0.21$. The full line is our theoretical prediction
for $\Gamma_{RW}$. The dashed line is an empirical 
expression that interpolates between the overdamped and underdamped limits 
of $\Gamma_{L}$ (Eq. 6.4 of Ref.\cite{hanggi_reaction-rate_1990}).
}
\label{fig:rate}
\end{figure}

Having determined $\Px(t_c)$ and $\Pu(t_c)$, we can compute
the escape rate $\Gamma_{RW}(t_c)$ at all $t_c$ through Eq.~\ref{eq:rate}.
The overdamped and the underdamped limits result
$\Gamma_{RW} = (1/2) \omega_b \pi^{-1} (\omega_0 t_c) e^{-\Delta U/T}$ and
$\Gamma_{RW} \propto t_c^{-1} e^{-\Delta U/T}$, respectively.
Fig.~\ref{fig:rate} shows that
our theoretical prediction (full line) well compares with numerical simulations 
of the model (open squares), at all $t_c$. 
In the figure, we also illustrate numerical results (full circles)
for the escape rate of the Langevin dynamics, $\Gamma_L$.
We remind that in the medium/high damping regime, and in the low temperature limit, $\Gamma_L$ is the celebrated Kramers' escape rate~\cite{kramers_brownian_1940},
$\Gamma_L = \frac{\omega_0}{2\pi}\left(\sqrt{1+\frac{\tau_{vis}^{-2}}{4\omega_b^2}}-\frac{\tau_{vis}^{-1}}{2\omega_b}\right) e^{-\Delta U/T}$,
and that finite temperature corrections have been determined by Lifson--Jackson~\cite{lifson_self-diffusion_1962}. 
In the overdamped limit, Kramer's result coincides with our prediction for $\Gamma_{RW}$ provided that 
$k = \omega_0/\omega_b$ in our functional form for $\Pu$.
In the underdamped limit, $\Gamma_L$ is known to scale as 
$\Gamma_{L} \propto \frac{\Delta U}{T} \tau_{\rm vis}^{-1} e^{-\Delta U/T}$ \cite{hanggi_reaction-rate_1990}.
This result clarifies that, as concern the escape rate, the two dynamics markedly differ in the underdamped
limit. The Langevin dynamics has a much higher escape rate, being $\Gamma_{L}/\Gamma_{RW} \propto \Delta U/T$.
 
The diffusion coefficient of the random walk dynamics is conveniently 
determined by describing each trajectory as a sequence of barrier crossing jumps, with displacement $\Dxe_i$,
each one followed by an effective intra-well jump, with displacement $\Dxi_i$.
The effective displacement $\Dxi_i$ is the total displacement of the intra--well
jumps connecting the final position of jump $\Dxe_i$, and the initial position of jump $\Dxe_{i+1}$.
Since the fraction of barrier-crossing jumps is $\Px$, after $N \simeq t/t_c$ jumps the overall 
displacement is $R_N = \sum_i^{N \Px} (\Dxe_i + \Dxi_i)$,
and the diffusion coefficient is
\begin{equation}
D = \lim_{t\rightarrow\infty}\frac{1}{2Nt_c}\left[\left(\sum_{j=1}^{N \Px}
\Dxe_j\right)^2
+\left(\sum_{j=1}^{N \Px}\Dxi_j\right)^2\right],
 \label{eq:divisione}
\end{equation}
%the cross product $\< \sum_i \Dxe_i \sum_j  \Dxi_j \>$ vanishing for symmetry reasons. 
the cross product term vanishing for symmetry reasons. 
Since only a fraction $\Pu$ of the terms appearing in the above sums are uncorrelated, 
the diffusion coefficient can also be expressed as
\begin{equation}
D = \frac{1}{2t_c}\Pu \Px \left[ \<(\Dxe)^2\> + \< (\Dxi)^2\> \right] = 
\De + \Di,
\label{eq:dshort}
\end{equation}
where $\<\cdot \>$ indicates {\it averages over uncorrelated jumps}.
Thus, we are left with the problem of estimating the mean square jump length
of uncorrelated barrier crossing jumps, $\<(\Dxe)^2\>$, and of uncorrelated
effective intra-well jumps, $\<(\Dxi)^2\>$. 

In the overdamped $t_c \to 0$ limit all jumps are short, and barrier crossing jumps
start and end close to the top of a barrier, where the potential is flat. 
Accordingly, $\< (\Dxe)^2\>_{t_c \to 0} \simeq 12 T t_c^2/m$ \footnote{The factor $12$ arises as one
is considering the mean squared length of free jumps crossing a fixed coordinate.}, and 
given our results for $\Pu$ and $\Px$, 
\begin{eqnarray}
\De_{t_c \to 0} \simeq & 6 T t_c \Px \Pu & \propto  T L^{-2} t_c^3 \Delta U e^{-\Delta U/T}. \label{eq:Deo}
\end{eqnarray}
To estimate $\Dxi_{t_c \to 0}$ we consider that,
since barrier crossing jumps are short, 
two subsequent barrier crossing jumps are connected by a sequence
of jumps whose total displacement is either zero, if the particle
recrosses the same barrier, or roughly equal to the period of the potential,
if the particle traverses a well and crosses a subsequent barrier.
For uncorrelated barrier crossing jumps these two 
possibilities are equally likely, which implies
$\< (\Dxi)^2\>_{t_c \to 0} \simeq L^2 / 2$. 
This allows to estimate
\begin{equation}
\Di_{t_c \to 0} \simeq \frac{L^2 }{4} \Px \frac{\Pu}{t_c} \propto \Delta U t_c  e^{-\Delta U/T}.  \label{eq:Dio}  
\end{equation}

In the underdamped $t_c \to \infty$ limit
a particle that has enough energy to cross an energy barrier will 
traverse $\Dxe(t,E)/L \simeq t/t_E$ wells, where $t$ is the jump duration,
and $t_E$ the time the particle needs to cross a single energy well.
Thus $\<(\Dxe(t,E))^2\> \simeq L^2 \<t^2\>/\<t_E^2\>$ 
is evaluated
averaging over the waiting time distribution ($\<t^2\> = 2 t_c^2$) and over the 
energy of the particle. This leads to
$\<(\Dxe(t,E))^2\> = 2t_c^2 L^2 \int_{\Delta U}^\infty \frac{e^{-\frac{E-\Delta U}{T}}}{t_E}dE \left( \int_{\Delta U}^\infty \hspace{-0.0cm} e^{-\frac{E-\Delta U}{T}} t_E dE \right)^{-1}$, 
that scales as
\begin{equation}
\<(\Dxe)^2\>_{t_c \to \infty} \propto \Delta U t_c^2
\label{eq:Dx_under}
\end{equation}
since $t_E(-\frac{L}{2} \to \frac{L}{2}) \propto \frac{L}{\sqrt{\Delta U}}$ for small $E-\Delta U$.
We thus estimate 
\begin{equation}
\De_{t_c \to \infty} \propto \Delta U t_c \exp(-\Delta U/T). \label{eq:Deu}
\end{equation}

To determine $\<(\Dxi)^2\>_{t_c \to \infty}$, we indicate with 
$p_k$ the probability that a walker interacts $k$ times with the heat bath in a well, before escaping. 
If $x_e$ is the original position inside the well, and $x_s$ the final one, then
$\<(\Dxi)^2\>_{t_c \to \infty}= \sum_k p_k \int P_{\rm e}(x_e) P_{\rm s}^{(k)}(x_s | x_e) (x_e - x_s)^2 dx_e dx_s$.
Here $P_{\rm e}(x) \propto 1/v(x)$ is the probability that a barrier crossing jump ends in $x_e$, one could evaluate
from the equilibrium distributions over the barrier-crossing jumps, and 
$P_{\rm s}^{(k)}(x_s | x_e)$ is the probability that the jump through which the particle escapes from the well 
starts in $x_s$, being the particle arrived in $x_e$.
To a good approximation~\cite{piscitelli_preparation_2016} a particle exits from the well after performing a single collision, so that $x_e-x_s = 0$,
or after thermalizing within the well, so that $x_e$ and $x_s$ becomes uncorrelated.
Accordingly, $\<(\Dxi)^2\>_{t_c \to \infty} = p_1 \cdot 0 + (1-p_1) \<\lambda^2\>$, where 
\begin{equation}
\<\lambda^2\>_{t_c \to \infty}\simeq \hspace{-0.15cm} \int_{-L/2}^{L/2} P_{\rm e}(x_e,T) P^{th}_{\rm s}(x_s,T)(x_e-x_s)^2 dx_e dx_s ,
 \label{eq:jumps}
\end{equation}
with $P^{th}_{\rm s}(x_s,T)$ the probability that a barrier crossing jumps of a thermalized state starts from position $x_s$.
The evaluation~\cite{piscitelli_preparation_2016} 
of both $p_1$, $P_{\rm e}(x_e,T)$ and $P^{th}_{\rm s}(x_s,T)$ leads to $\<(\Dxi)^2\>_{t_c \to \infty} \propto L^2$.  
Summarizing, in the $t_c \to \infty$ limit 
\begin{equation}
\Di_{t_c \to \infty} = \frac{\Px}{2t_c} \<(\Dxi)^2\>_{t_c \to \infty}  \propto \frac{L^2}{t_c} \exp(-\Delta U/T) \label{eq:Diu}.
\end{equation}
We finally note that, in both Eq.~\ref{eq:Deu} and Eq.~\ref{eq:Diu}, the proportionality constants have a weak temperature dependence
we neglect, that is fixed by the shape of the potential~\cite{piscitelli_preparation_2016}.

\begin{figure}[t]
\centering
\includegraphics*[scale=0.33]{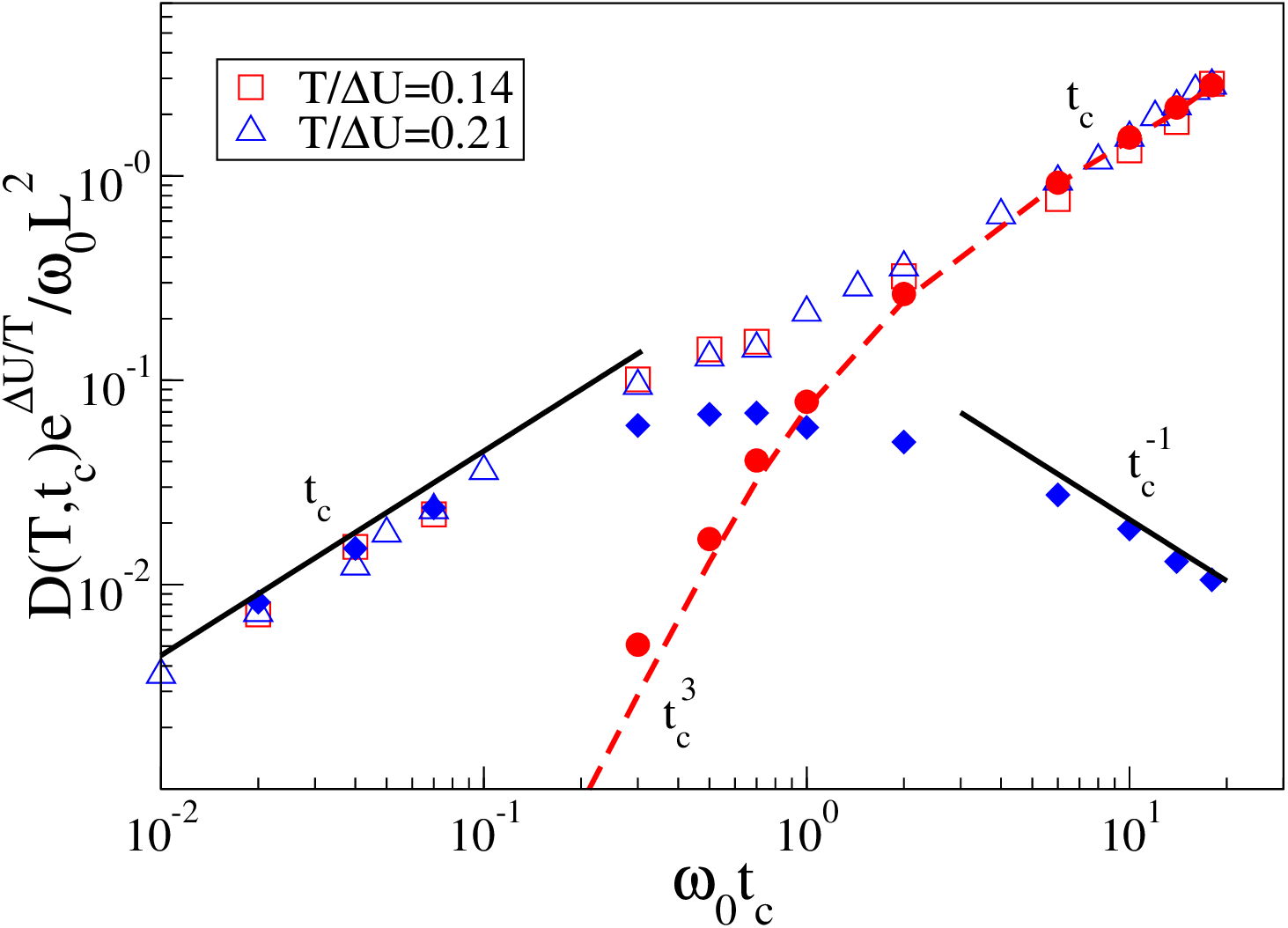}
\caption{Open symbols illustrate the dependence of the normalized diffusion coefficient on the typical
time of collision with the heat bath at low temperatures. Full symbols
indicate the contribution due to $\Di$ (diamonds) and $\De$ (circles). 
Full lines are predictions
for $\Di$ in the overdamped and in the underdamped limits, while the dashed line
is the analytical prediction of $\De$, at all $t_c$. All predictions are for $T/\Delta U = 0.21$.
}
\label{fig:fig2}
\end{figure}

While we have estimated $\De_{t_c \to 0}$ and $\De_{t_c \to \infty}$ in the low temperature regime, 
it is also possible to estimate $\De(t_c)$ at all $t_c$. To this end we assume 
the barrier crossing jumps to be always uncorrelated, since they 
are uncorrelated in the overdamped limit, as jumps are short and 
particles on the top of the barrier move as free particles, and in the underdamped limit.
With this assumption we estimate $\Px \<(\Dxe)^2\>$ from  equilibrium average over the jumps,
$\Px \<(\Dxe)^2\> =2 \int_{-L/2}^{L/2} dx_s \int_{L/2}^\infty dx_e \int_{0}^\infty dE f(x_s,x_e, E)(x_e-x_s)^2$, 
with $f$ given in Eq.~\ref{eq:f}, and thus get $\De(t_c) = \frac{1}{2t_c}\Px \Pu \<(\Dxe)^2\>$.
Beside being valid at all $t_c$ in the low temperature regime, this prediction is also valid at all
temperature in the underdamped regime, where $\Pu = 1$.
Fig.~\ref{fig:fig2} illustrates that this theoretical prediction (dashed line)
correctly describes the numerical data (full circles), and scales as $t_c^3$ and as $t_c$ in the overdamped
and in the underdamped limit, as predicted in Eq.~\ref{eq:Deo} and in Eq.~\ref{eq:Deu}, respectively.
In the figure, we also illustrate numerical results for the contribution to the diffusion
coefficient of the intra-well jumps (full diamonds), that behaves as predicted 
in Eq.~\ref{eq:Dio} and in Eq.~\ref{eq:Diu} in two limits.
Thus, the overall diffusion coefficient exhibits
a crossover between two linear regimes, as $D \simeq \Di$ in the overdamped limit,
and $D \simeq \De$ in the underdamped one.

\begin{figure}[t]
 \centering
 \includegraphics[scale=0.33]{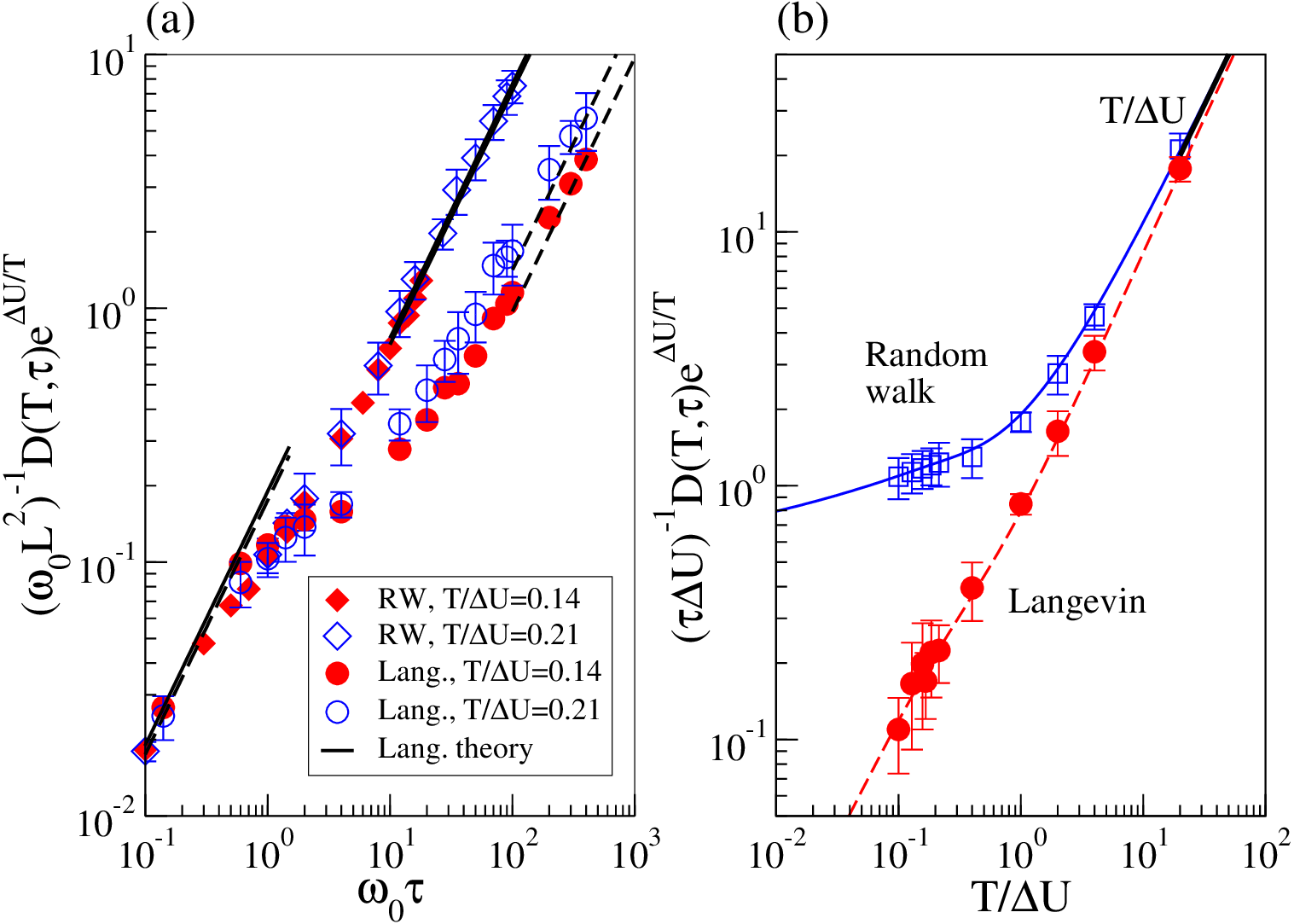}
 \caption{
 Diffusivities of the Random Walk and of the Langevin dynamics, assuming $\tau = \tau_{\rm vis} = t_c$.
 Panel (a) illustrates the $\tau$ dependence of the diffusivities, for different temperatures, while panel (b)
 illustrates the dependence of the diffusivity on the temperature, in the underdamped regime ($\tau \omega_0 = 10^2$).
 The full line represents Eq.~\ref{eq:Deu} for the Random Walk dynamics and the dashed line represents
 Eq.~3.9 of Ref.\cite{pavliotis_diffusive_2008} for the Langevin dynamics.}
 \label{fig:fig3}
 \end{figure}

We finally compare the diffusion coefficient of the random walk and of the Langevin dynamics,
identifying their characteristic timescales, $\tau = t_c = \tau_{\rm vis}$.
For both dynamics $D = \Gamma L^2$ in the overdamped limit. In this limit,
the full van Hove distributions actually coincide at all times.
In the underdamped low temperature limit, the diffusivity of the Langevin dynamics is
$D_L \propto T\tau e^{-\Delta U/T}$~\cite{pavliotis_diffusive_2008},
while that of the random walker is given by Eq.~\ref{eq:Deu}, $D \propto \Delta U \tau e^{-\Delta U/T}$. 
Accordingly, in this limit the random walk dynamics is faster than the Langevin one, as illustrated in Fig.~\ref{fig:fig3}a.
Fig.~\ref{fig:fig3}b compares the diffusivities of the two dynamics as concern their temperature dependence, in the underdamped regime.
The numerical results for the temperature dependence of the random walk diffusivity are correctly described by our theoretical prediction 
for $\De$ valid at all temperatures (full line), while those of the Langevin dynamics have been predicted in Ref.~\cite{pavliotis_diffusive_2008}.
At high temperature, $T > \Delta U$, the effect of the potential is negligible and the two diffusivities
coincide, and scale as $\tau T e^{-\Delta U/T}$. In the low temperature regime, the Langevin diffusivity
does not change temperature dependence, while the diffusivity of the random walk model only depends on temperature through the Arrhenius factor.
The difference in the diffusivities in the underdamped low temperature regime is rationalized considering that 
the two dynamics are mapped on free random walks with a different jump rate $\Gamma$, and a different mean square jump length, $\<\lambda^2\>$. 
Indeed, on the one side we have already seen that $\Gamma_{L}/\Gamma_{RW} \propto \Delta U/T$.
On the other side, in the underdamped limit the mean square size of the jumps of the Langevin dynamics~\cite{pavliotis_diffusive_2008} 
scales as $\<\lambda^2\>_L \propto T^2\tau_{\rm vis}^2 / \Delta U$, while that of the random walk dynamics
scales as $\<\lambda^2\>_{RW} \propto \Delta U t_c^2$, as in Eq.~\ref{eq:Dx_under},
so that $\<\lambda^2\>_{L}/\<\lambda^2\>_{RW} \propto (T/\Delta U)^2$. Thus, despite making more frequent irreversible barrier
crossing jumps, the Langevin dynamics has a smaller diffusivity as its jumps are much shorter.

In conclusion, we put forward an analytical treatment of the escape rate from a well, and of the
diffusion coefficient in a periodic potential, of a random walker, in both the overdamped
and the underdamped limits. The walker behaves as a Langevin particle in the overdamped
limit. In the underdamped low temperature limit, conversely, with respect to a Langevin particle
a random walker has a smaller
escape rate, but a larger diffusion coefficient. An important open question ahead concerns the 
temporal evolution of the van Hove distribution in the underdamped limit, 
where a transient Fickian not-Gaussian dynamics is observed, as
in many physical systems~\cite{wang_when_2012,kindermann_nonergodic_2016}.

\begin{acknowledgments}
Support from the Singapore Ministry of Education through the Academic Research Fund (Tier 1) under Projects No. RG104/15 and RG179/15 is gratefully acknowledged.
\end{acknowledgments}

%\bibliography{RandomWalkPotential} 
%

\end{document}